\documentclass[journal=jctcce,manuscript=article]{achemso}

\usepackage[utf8]{inputenc}
\usepackage{amsmath,amssymb,pdfpages,geometry,graphicx,placeins,enumitem,booktabs} 
\usepackage[hidelinks]{hyperref}
\usepackage[labelfont=bf]{caption}
\usepackage{microtype}
\usepackage{setspace}

\makeatletter
\patchcmd{\env@cases}{1.2}{0.6}{}{}
\makeatother

\renewcommand{\vec}[1]{\mathrm{\mathbf{#1}}}

\AddToHook{cmd/cite/before}{\ifhmode\unskip~\fi}

\title{Structure determination from single-molecule X-ray scattering images using stochastic gradient ascent.}

\author{Steffen Schultze}
\affiliation{Max Planck Institute for Multidisciplinary Sciences}
\author{D.\ Russell Luke}
\affiliation{Institute for Numerical and Applied Mathematics, University of Göttingen}
\author{Helmut Grubmüller}
\email{hgrubmu@mpinat.mpg.de}
\affiliation{Max Planck Institute for Multidisciplinary Sciences}

\begin{document}
% \onehalfspacing
\setlength\itemsep{0.5ex}

\maketitle 

\begin{abstract}
Scattering experiments using ultrashort X-ray free electron laser (XFEL) pulses have opened a new path for structure determination of a wide variety of specimens, including nano-crystals and entire viruses, approaching atomistic spatial and femtoseconds time resolution. However, random and unknown sample orientations as well as low signal to noise ratios have so far prevented a successful application to smaller specimens like single biomolecules.
We here present resolution-annealed stochastic gradient ascent (RASTA), a new approach for direct atomistic electron density determination, which utilizes our recently developed rigorous Bayesian treatment of single-particle X-ray scattering. We demonstrate electron density determination at $2\text{\AA}$ resolution of various small proteins from synthetic scattering images with as low as $15$ photons per image.
 
\end{abstract}

\clearpage

\section{Introduction}

\begin{figure}[ht]
    \includegraphics[trim={0 20px 0 150px},clip,width=\textwidth]{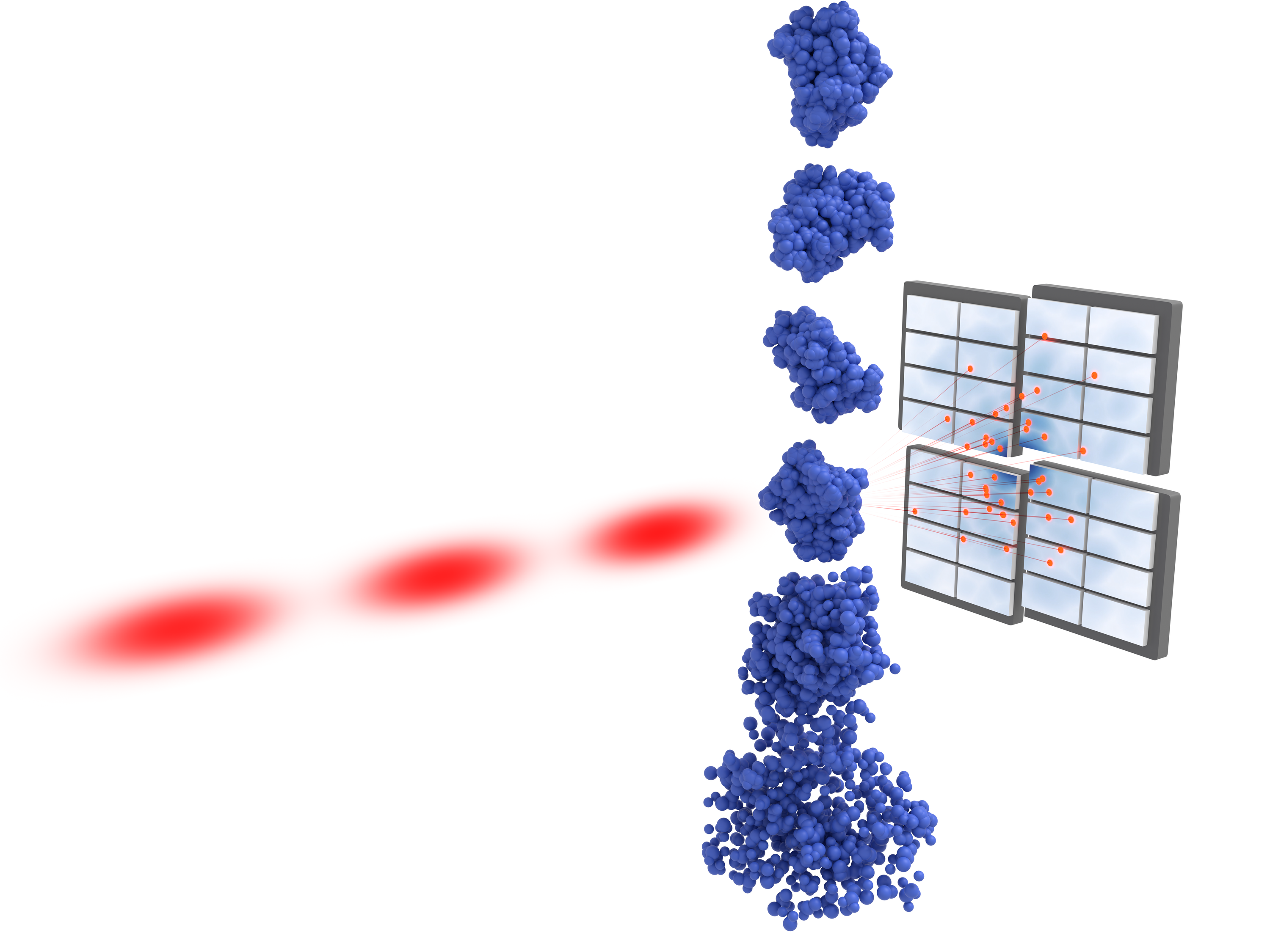}
    \caption{Single molecule scattering experiment. }
    \label{fig: introexperiment}
\end{figure}

Single-particle X-ray scattering experiments using ultrashort X-ray free electron laser pulses (XFELs) allow for electron density determination approaching atomistic spatial and femtoseconds time resolution \cite{hajdu_single-molecule_2000, huldt_diffraction_2003, gaffney_imaging_2007, miao_beyond_2015}.
In these `diffraction before destruction' \cite{chapman_diffraction_2014} experiments (Fig.~\ref{fig: introexperiment}), each sample particle is exposed to a high-intensity X-ray pulse, and the resulting scattering image is recorded on a detector. 
The femtosecond pulse duration ensures that the scattering outruns the destruction of the sample, which therefore is observed essentially without radiation damage \cite{neutze_potential_2000}.

Many specialized approaches for structure determination from these experiments have been developed, including approaches for orientation determination \cite{shneerson_crystallography_2008, loh_reconstruction_2009, walczak_bayesian_2014, kassemeyer_optimal_2013, elser_three-dimensional_2011, tegze_atomic_2012, flamant_expansion-maximization-compression_2016, ayyer_dragonfly_2016}, manifold embedding \cite{fung_structure_2009,schwander_symmetries_2012, giannakis_symmetries_2012, winter_enhancing_2016}, for phase retrieval \cite{elser_searching_2007, luke_relaxed_2004}, or using photon correlations \cite{saldin_structure_2009, saldin_reconstructing_2011, saldin_beyond_2010, saldin_new_2011, saldin_structure_2010, starodub_single-particle_2012, kurta_correlations_2017, donatelli_iterative_2015, von_ardenne_structure_2018}, and have been successfully applied to a broad variety of specimens, including nano-crystals \cite{schlichting_serial_2015, chapman_structure_2017,oda_time-resolved_2021,tenboer_time-resolved_2014} and entire viruses \cite{seibert_single_2011, ekeberg_three-dimensional_2015, hosseinizadeh_high-resolution_2014}.
However, while the idea of performing these experiments also on single macromolecules was proposed over two decades ago, their relatively small size presents formidable challenges, due to the much smaller expected number of typically $10$ to $100$ scattered photons per image \cite{von_ardenne_structure_2018}.
As a consequence of these low photon counts, and in contrast to larger samples, for small proteins it is fundamentally impossible to determine their orientation, which is random and unknown for each image. 
Most approaches however rely on this orientation information, or, in the case of photon correlations \cite{saldin_structure_2009,von_ardenne_structure_2018}, discard too much information to be applicable at realistic numbers of images. 

To circumvent this problem, we have recently developed a rigorous Bayesian treatment \cite{schultze_bayesian_2024} of single-particle X-ray scattering. Rather than attempting to determine the sample orientation for each scattering image, this approach rests on a likelihood function for the full set of scattering images that includes a probability weighted average over all possible molecular orientations. 
In addition, it allows for systematic inclusion of experimental noise and uncertainties such as background noise, polarization, and incomplete detector coverage. 
As a further benefit, the electron density is directly determined in real space, thus also circumventing the phase problem. 
In contrast to orientation determination methods, however, our Bayesian approach does not directly construct an electron density map. 
Instead, it requires optimizing or sampling from the Bayesian posterior using techniques like Markov chain Monte Carlo (MCMC).
As is typical for MCMC, the computational cost increases rapidly for more complex structures, which has so far limited the application of this approach to rather small structures with a few hundred degrees of freedom. 

We here present a new technique for efficient calculation of the gradient of the logarithmic posterior, and a new optimization method we call resolution-annealed stochastic gradient ascent (RASTA), which, as we will show, reduces the computational cost of the optimization by a factor of up to $1000$. We assess accuracy and efficiency of this approach on a number of synthetic test cases, including the direct determination of all $1300$ atomic positions of the protein Lysozyme \cite{kuroki_covalent_1993-1} from only $10^6$ simulated scattering images.

\section{Theory and Methods}

\subsection{Bayesian formalism}
We seek to determine the electron density $\rho$ that maximizes the Bayesian posterior probability $P(\rho\,|\,\mathcal I)$ of the molecular electron density $\rho$ given the \emph{full} set of observed scattering images $\mathcal I$.
By Bayes' theorem this posterior is given by the likelihood $P(\mathcal I\,|\,\rho)$ of the images given $\rho$ and the prior $P(\rho)$,
\begin{equation}
    P(\rho\,|\,\mathcal I) \propto P(\mathcal I \,|\, \rho) P(\rho).
\end{equation}
The likelihood, in turn, decomposes into a product over all $N$ single images,
\begin{equation}
    P(\mathcal I \,|\, \rho) = \prod_{j=1}^N P(\vec k_1^{(j)}, \dots, \vec k_{n_j}^{(j)} \,|\, \rho).
\end{equation}
Here, for each scattering event $j = 1\ldots N$, the resulting scattering image is described by the positions of the $n_j$ scattered photons on the detector, given by their scattering vectors $\vec k_1^{(j)}, \dotsc, \vec k_{n_j}^{(j)}$. 
The joint distribution of these scattering vectors is described by the single-image likelihood \cite{schultze_bayesian_2024}
\begin{equation}\begin{split} \label{eq: single-image likelihood}
        & P(\vec k_1, \dots, \vec k_n \,|\, \rho) = \int_{\mathrm{SO}(3)} \exp\left(-I_0 \int_E \lvert \hat{\rho}(\vec R\vec k)\rvert^2 \,\mathrm{d}\vec k\right)\prod_{i=1}^n \lvert \hat{\rho}(\vec R\vec k_i)\rvert^2  \,\, \mathrm{d}\vec R,
\end{split}\end{equation}
where $I_0$ is the incoming beam intensity and $E$ denotes the Ewald sphere, which implements the forward model specified by the physics of the scattering event.
We here focus on a background-free forward model, including only photons from coherent scattering described by the Fourier transform $\hat\rho$ of the electron density. 
The unknown molecular orientation, described by a rotation matrix $\vec R$, is marginalized out by integrating over the rotation group $\mathrm{SO}(3)$.

Whereas the Bayesian formalism as presented so far is independent of a specific representation of the electron density $\rho$, the prior $P(\rho)$ is inherently dependent on the atomic positions. Therefore, we use a representation of $\rho$ by a linear combination of Gaussian beads,
\begin{equation}\label{eq: representation}
    \rho(\vec r) = \sum_{i=1}^m \frac{h_i}{\left(w_i\sqrt{2\pi}\right)^{\mkern-2mu 3}} \exp \left(-\frac1{2w_i^2}\lVert \vec r - \vec y_i\rVert^2\right),
\end{equation}
with bead positions $\vec y_i$, heights $h_i$ and widths $w_i$. These beads may both represent single atoms or describe larger structural entities. 
At the targeted atomistic resolution level one common height $h = h_i$ and one common width $w = w_i$ is sufficient to represent structures by specifying the positions of the heavy atoms and neglecting the hydrogen atoms.  
For ease of notation we will omit these parameters in the following, and write $\rho_{\vec y_i}(\vec r)$ for the corresponding electron density.

We include two kinds of prior information on the bead positions $\vec y_i$. First, they should not occupy the same position, and second, they should form one connected molecule. 
Including both of these restrictions, the prior reads
\begin{equation}
    -\log P(\rho_{\vec y_i}) = \sum_{i=1}^m \sum_{j=1}^m f_1(\lVert \vec y_i - \vec y_j \rVert) + \sum_{i=1}^m f_3\left(\sum_{j=1}^m f_2(\lVert \vec y_i - \vec y_j \rVert)\right),
\end{equation}
where $f_1(\lVert \vec y_i - \vec y_j \rVert)$ is a short range repulsive pair potential, the term $\sum_{j=1}^m f_2(\lVert \vec y_i - \vec y_j \rVert)$ smoothly counts the number of neighbors of each bead within a certain radius, and $f_3$ is a potential ensuring that this number of neighbors remains above a certain threshold. We refer to the supplement for the exact functional forms of $f_1$, $f_2$ and $f_3$.

\subsection{Computational aspects}
\begin{figure}[ht]
    \centering
    \includegraphics[width=0.66\textwidth]{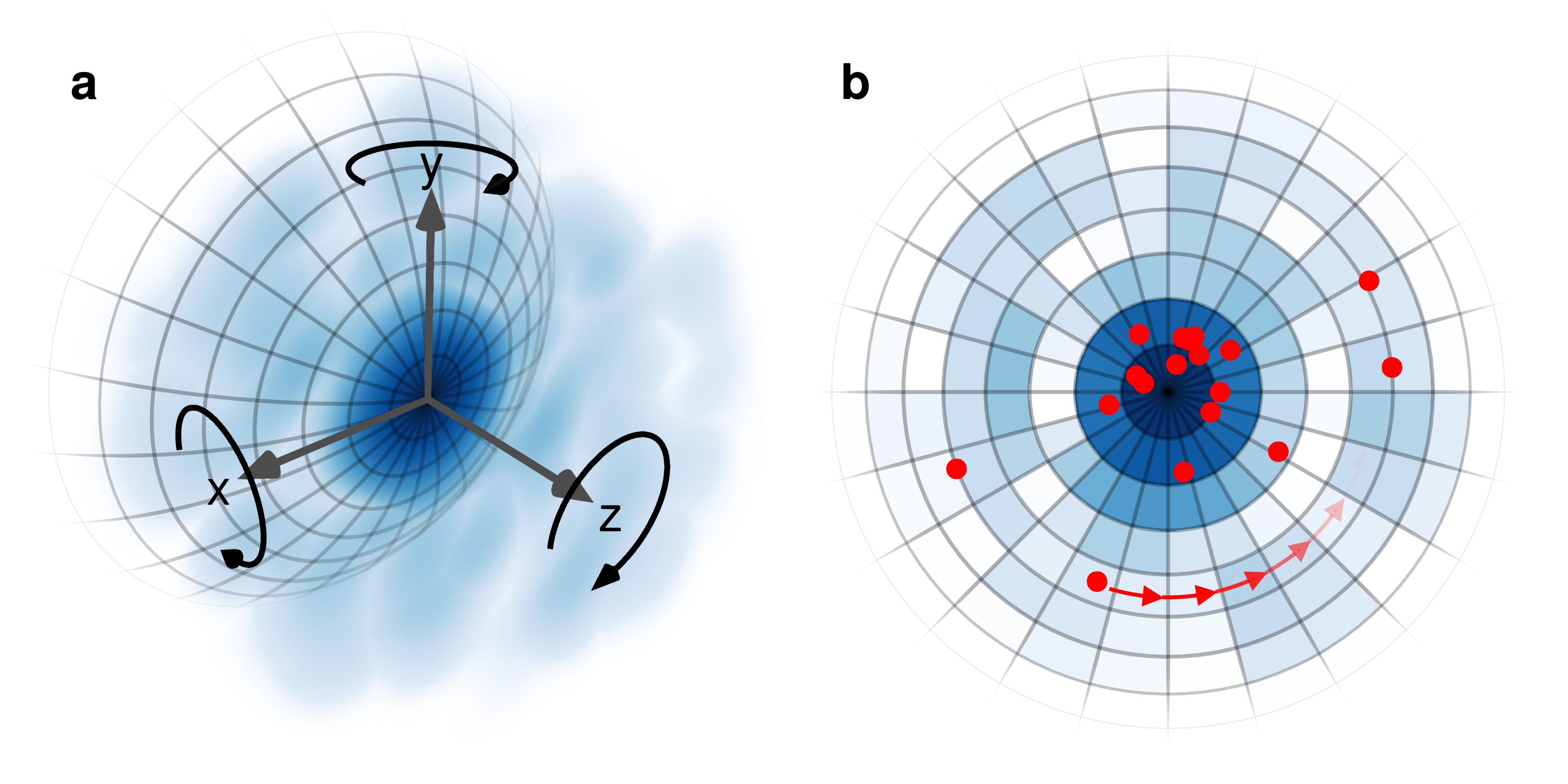}
    \caption{Computation of likelihood. \textbf{a} Exemplary orientation of the Ewald sphere (mesh) within the Fourier intensity function $I(\vec k) = \lvert \hat\rho(\vec k)\rvert^2$ (blue density). \textbf{b} Rotations around the beam axis correspond to rotations of the photon positions (red dots) within the intensity maps in polar coordinates (blue density). Shown grid dimensions are reduced for visual clarity.}
    \label{fig: computation}
\end{figure}

Our optimization approach relies on efficient computations of the gradient of the log-likelihood $\log P(\mathcal I\,|\,\rho)$. 
To this end, first consider the computation of the likelihood \cite{schultze_bayesian_2024}.
The integral over $\vec R$ in the single-image likelihood from eq.~\eqref{eq: single-image likelihood} is approximated in two steps, as illustrated in Fig.~\ref{fig: computation}. 
First, the degrees of freedom around the $x$ and $y$ axes are taken into account via a weighted average over orientations $\vec R_l$ of the Ewald sphere with weights $w_l$ for $1 \leq l \leq n_\vec{R}$ (Fig.~\ref{fig: computation}a). 
Second, for the $z$-axis degree of freedom (Fig.~\ref{fig: computation}b), corresponding to rotations around the beam axis, the intensity function is evaluated on a polar coordinate grid (blue densities in Fig.~\ref{fig: computation}b) for each $l$, setting $I_{l,r,s} = \lvert\hat\rho(\vec R_l \vec q_{r,s})\rvert^2$ where $\vec q_{r,s}$ are the center positions of the grid cells, indexed by $1 \leq r \leq n_r$ and $1 \leq s \leq n_s$ in the radial and angular direction, respectively. 
For each photon position $\vec k_i$ (red dots in Fig.~\ref{fig: computation}b), let $r(\vec k_i)$ and $s(\vec k_i)$ denote the radial and angular index of the corresponding grid cell, respectively. 
The rotations around the beam axis are then equivalently described as offsets $s'$ of the angular grid indices $s(\vec k_i)$, and are therefore included by summing over all these offsets.
Including both of these steps, the approximate single-image likelihood now reads
\begin{equation}\label{eq: likelihood approx}
    P(\vec k_1, \dots, \vec k_n \,|\, \rho_{\vec y_i}) \approx \sum_{l = 1}^{n_\vec{R}} w_l \exp(-I_0 \lambda_l) \frac1{n_s} \sum_{s'=1}^{n_s} \prod_{i=1}^n I_{l,r(\vec k_i),s(\vec k_i)+s' \bmod {n_s}}.
\end{equation}
Here, the integral over the Ewald sphere $E$ in eq.~\eqref{eq: single-image likelihood}, which only depends on $\vec R_l$, is estimated from the same intensity grids as
\begin{equation}\label{eq: numerical nphotons}
    \lambda_l = \sum_{r=1}^{n_r} \sum_{s=1}^{n_s} a_{r,s} I_{l,r,s},
\end{equation}
where $a_{r,s}$ is the area of each grid cells. As a further benefit of the polar intensity maps, the density of grid cells is naturally higher near the origin where the intensity is highest, improving the accuracy of this estimate. 

From the above approximation of the likelihood, the gradient is calculated by backpropagation, that is, by first computing the derivatives with respect to each $I_{l,r,s}$ and then applying the chain rule to obtain the gradients with respect to $\vec y_i$, as detailed in the supplement. 
Both the computation of the likelihood in eq.~\eqref{eq: likelihood approx} as well as of the gradient were implemented as CUDA-kernels in the Julia programming language \cite{bezanson_julia_2017}, which we provide as part of our open source software at \url{https://gitlab.gwdg.de/sschult/xfel}.
Utilizing the special treatment of the beam axis rotations as explained above is essential for performance, reducing both the memory requirement as well as the computational effort by a factor of over $100$. 

\subsection{Resolution-annealed stochastic gradient ascent}
A naive application of gradient ascent on the log-posterior $-\log P(\rho_{\vec y_i}\,|\,\mathcal I)$ would fail, as it is highly non-convex and has exceedingly many local minima. The reason for this lies in the Fourier-nature of the scattering images, with photons at low-$k$ corresponding to information about large length scales, and photons at high-$k$ to small scale information. 
Consequently, each possible electron density is surrounded by local minima induced by high-$k$ contributions. % check if this is actually correct

\begin{figure}[ht]
    \centering
    \includegraphics[width=1\linewidth]{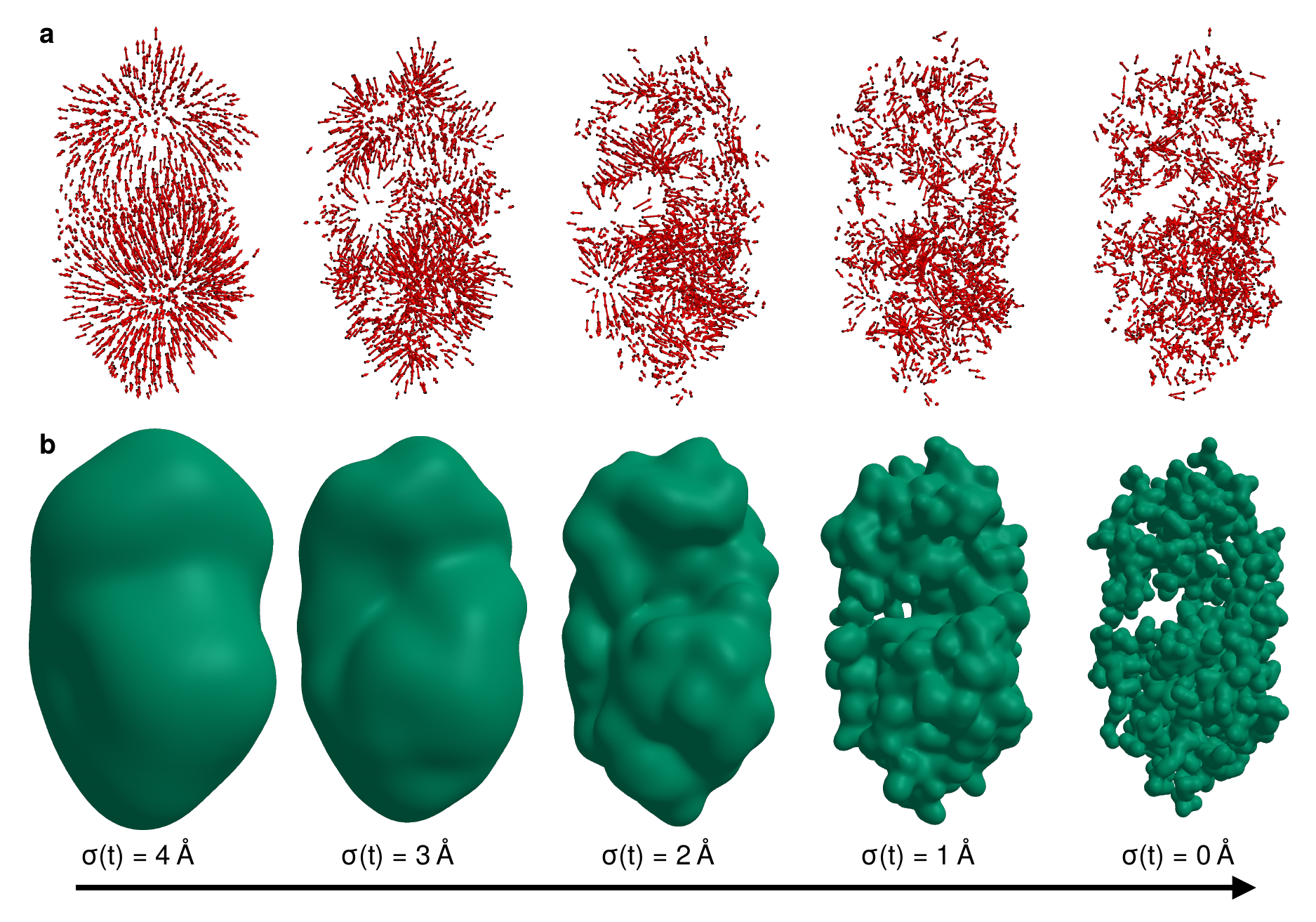}
    \caption{Snapshots from an exemplary optimization run (on the Lysozyme, PDB 148L), at optimization times $t$ where $\sigma(t)$ reaches selected thresholds. \textbf{a} Gaussian bead positions (black dots) and current stochastic gradient (red arrows). \textbf{b} Electron density snapshots.}
    \label{fig: optimization}
\end{figure}

We therefore developed a `resolution-annealed stochastic gradient ascent' (RASTA), which circumvents these minima by removing the high-$k$ photons in the beginning of the optimization, and slowly adding them back in over the optimization steps $t$. 
To this end, at each optimization step $t$ we consider a smoothed version $S_{\sigma(t)}(\rho_{\vec y_i}) = \rho_{\vec y_i} * \mathcal N(0, \sigma(t))$ of the electron density obtained by convolution with a Gaussian kernel $\mathcal N(0, \sigma(t))$, and exploit the Fourier convolution theorem to obtain a set $S_{\sigma(t)}(\mathcal I)$ of scattering images corresponding to this smoothed density by rejection sampling. 
Each update is then performed according to the gradient of the log-posterior of $S_{\sigma(t)}(\rho_{\vec y_i})$ given $S_{\sigma(t)}(\mathcal I_t)$,
\begin{align}
    \vec v_i(t + 1) &= \beta(t) \vec v_i(t) + \frac{\partial}{\partial \vec y_i} \log P \left(S_{\sigma(t)}(\mathcal I_t) \,\middle|\, S_{\sigma(t)}(\rho_{\vec y_i(t)})\right), \\
    \vec y_i(t + 1) &= \vec y_i(t) + \eta(t) \vec v_i(t + 1),
\end{align}
where $\mathcal I_t$ denotes a random batch of images used for step $t$.
The effect of this smoothing on the gradient can be seen in Figure~\ref{fig: optimization}, which shows an exemplary optimization run on the Lysozyme protein (as explained in the results section). 
As can be seen, the smoothed gradient (Fig.~\ref{fig: optimization}a) corresponds to large-scale of the smoothed electron density (Fig.~\ref{fig: optimization}b) updates during the early phase of the optimization and subsequently smaller scale updates as the optimization proceeds. 
The smoothing scale $\sigma(t)$ is reduced to zero with increasing $t$, at which point all photons are included. 
Also the step size $\eta(t)$ and the momentum parameter $\beta(t)$ are properly adapted during the optimization, as described in the supplement. 

This approach is stochastic not only due to the random batches $\mathcal I_t$ of images used for each step \cite{luke_stochastic_2024}, but also because the rejection sampling to obtain $S_{\sigma(t)}(\mathcal I)$ has a different random outcome for each step. 
As a result, although the rejection sampling discards many photons for each single step, with increasing iterations each photon will be included almost surely, such that no information is lost even in the early stages of optimization.

\FloatBarrier
\clearpage
\section{Results and Discussion}
\begin{figure}[ht]
    \centering
    \includegraphics[width=1\linewidth]{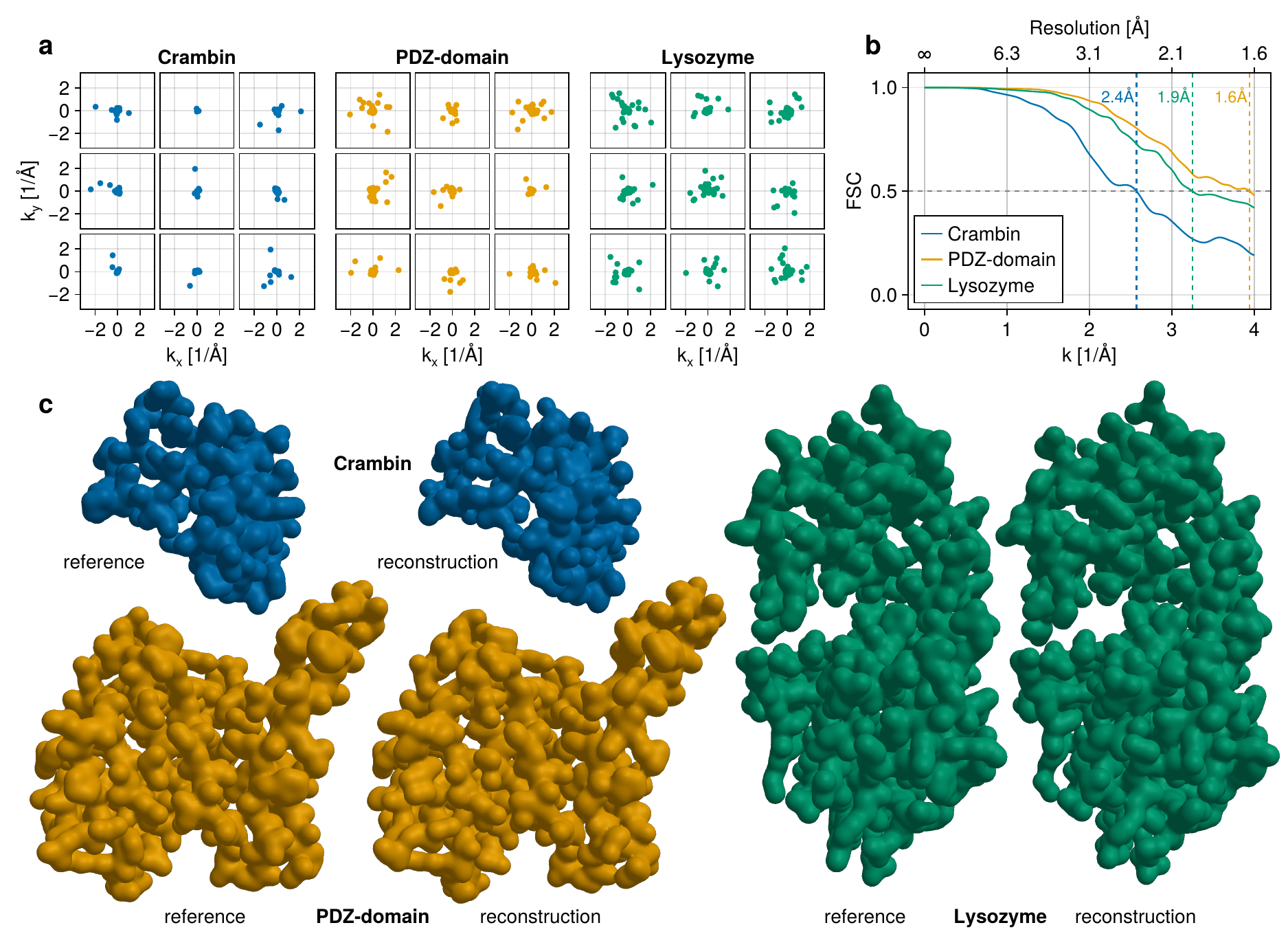}
    \caption{Exemplary application on Crambin, PDZ-domain, and Lysozyme. \textbf{a} Exemplary simulated scattering images. Note that the majority photons is scattered to low $k$. \textbf{b} Fourier shell correlations show achieved resolutions of around $2\,\text{\AA}$. \textbf{c} Comparison of reference and reconstructed electron densities.}
    \label{fig:examplereconstructions}
\end{figure}

To assess the accuracy, achievable resolution, and computational efficiency of our approach, we selected three small to medium-sized globular proteins as test cases (Table~\ref{tab: parameters}), the $46$-residue protein Crambin (PDB 1EJG \cite{jelsch_accurate_2000}), the $112$-residue neuronal nitric oxide synthase PDZ-domain (PDB 1QAU \cite{hillier_unexpected_1999}), and the $167$-residue protein Lysozyme (PDB 148L \cite{kuroki_covalent_1993-1}). 
For each case, between $10^6$ and $10^7$ synthetic scattering images were generated as described in our previous study \cite{schultze_bayesian_2024} from the reference structure taken from the protein data bank \cite{berman_protein_2000}. 
Examples of these scattering images are shown in Fig.~\ref{fig:examplereconstructions}a.
All reconstructions were performed using one Gaussian bead per heavy atom in the reference structure. 
Detailed information on optimization parameters, annealing, and step size schedules is provided in the the supplement. 
All three runs reached reconstructions with a likelihood greater than that of the corresponding reference structures, indicating that the global maxima were reached.

\begin{table}[ht]
    \begin{tabular}{lllllll}
        \toprule
        Name & PDB & \shortstack[l]{heavy \\ atoms} & \shortstack[l]{expected photons \\ per image} & \shortstack[l]{number of\vphantom{p} \\ images} \\% & batch size & \shortstack[l]{number of\vphantom{p} \\ SGD steps} \\
        \midrule
        Crambin & 1EJG & $327$ & $15$ & $10^7$ \\% & $200\,000$  & $50\,000$\\
        PDZ-domain & 1QAU & $812$ & $42$ & $10^6$\\% & $50\,000$  & $20\,000$ \\
        Lysozyme & 148L & $1300$ & $79$ & $10^6$ \\% & $50\,000$ & $20\,000$\\
        \bottomrule
    \end{tabular}
    \caption{Test cases}
    \label{tab: parameters}
\end{table}

Figure~\ref{fig:examplereconstructions}c compares the reference structures with the obtained reconstructions. As can be seen, the atomistic positions are nearly completely recovered, as is corroborated by obtained Fourier-shell-correlation (FSC) resolutions of $2\,\text{\AA}$ and even higher (Fig.~\ref{fig:examplereconstructions}b). 
For this resolution estimate, a conservative cutoff of $0.5$ was used, using other popular cutoffs the resolutions would be even higher \cite{van_heel_fourier_2005}.
As a further measure of quality, the optimal transport plans between the bead positions of reconstructed and reference structures were computed, obtaining earth-mover's distances of $0.9\,\text{\AA}$ for Crambin, $0.6\,\text{\AA}$ for the PDZ-domain, and $0.65\,\text{\AA}$ for Lysozyme. 

\begin{figure}[ht]
    \centering
    \includegraphics[width=1\linewidth]{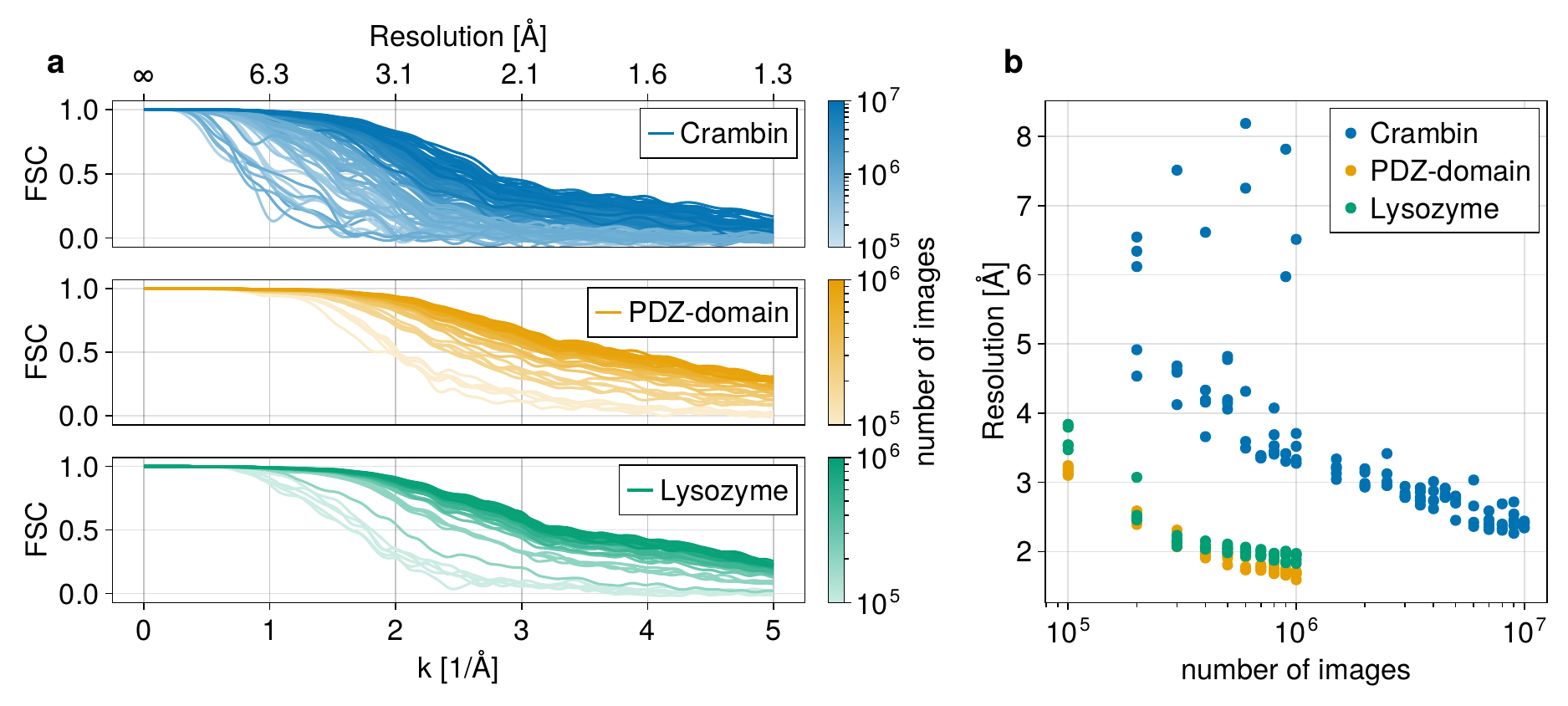}
    \caption{Resolution as a function of the number of images. \textbf{a} Fourier shell correlations between reference structure and reconstruction for each test run. \textbf{b} Achieved FSC-resolution for each run.}
    \label{fig: scaling}
\end{figure}

Interestingly, despite the $10$-times higher number of images, the achieved resolution is lowest for the smallest test protein Crambin. This finding is, at first glance, quite counterintuitive, as one might expect that the reduced number of unknowns would require less data. 
However, for smaller specimens the images contain fewer photons, and therefore less information. 
To further investigate this information content, we next asked how the achieved resolution depends on the number of images. 
To answer this question, we performed a number of independent optimization runs for each test case using varying numbers of images (Fig.~\ref{fig: scaling}). Figure~\ref{fig: scaling}a shows the Fourier-shell-correlation for each run, and Figure~\ref{fig: scaling}b the FSC-resolution obtained at the conservative threshold of $0.5$. 

Indeed, for the smallest protein (Crambin) far more images are required to achieve a similar resolution than for the others. For instance, to reliably achieve $2.5\,\text{\AA}$ resolution, $10^7$ images are required for Crambin, while already $2\cdot 10^5$ images suffice for the other test proteins. 
These values differ by a factor of $50$, which is much higher than the difference in photon counts per image, showing that the information content per image decreases more than linearly with the number of photons. 

It is further notable that the required number of images does not appear to strictly decrease with the number of atoms, either. 
Indeed, using the same number of images higher resolution is obtained for the PDZ-domain than for the larger Lysozyme. We assume this inversion is due to the less symmetric shape of the PDZ-domain. 

\FloatBarrier

\section{Conclusions}
We have developed and assessed a resolution-annealed optimization approach using stochastic gradient descent to determine electron densities at near-atomistic resolution from sparse single-molecule X-ray scattering images. 

Compared to previous approaches, the computational effort to achieve high resolution was reduced drastically, making structure determination of small to medium-sized proteins and other biomolecules computationally feasible.
Not only is the required compute time highly reduced for any given resolution, but our approach also should also enable one to achieve resolutions that were previously unattainable. 
For instance, our previous hierarchical MCMC approach \cite{schultze_bayesian_2024} reached computational limits already at $4\,\text{\AA}$ resolution, requiring more than $1000$ GPU-hours in total. 
In comparison, the implementation of the approach developed here achieves $2\,\text{\AA}$ requiring only a few GPU-hours, and $4\,\text{\AA}$ within a few minutes.
Further, the computational effort of the approach presented here is independent of the number of images, and only directly depends on the chosen batch size, which will be pivotal for future application to noisy experimental data.

Generalizing this optimization approach to images with background photons in addition to the coherently scattered photons included here, as we have already specified \cite{schultze_bayesian_2024}, is straightforward, although substantially more images will be required. 
Importantly, our Bayesian approach is also very efficient in terms of the required number of images; in fact, it utilizes the complete available structural information and is therefore in this regard optimal. 
The comparison to approaches based on photon correlations, which are the only other methods applicable in this extreme Poisson regime, is particularly striking. 
For instance, whereas the three-photon correlation method \cite{von_ardenne_structure_2018} required about $2\cdot 10^9$ images to achieve a $3.3\,\text{\AA}$ resolution for Crambin, our tests demonstrated that only $1.5\cdot 10^6$ suffice to achieve similar resolution. 
This over thousand-fold improvement will likely translate to realistic images with background noise.

From a more general viewpoint, cryogenic electron microscopy (cryo-EM) shares many similarities with single-molecule X-ray scattering, including random and unknown molecule orientations. 
Although our rejection-sampling based smoothing approach does not have an exact analogue, an similar resolution-annealed stochastic gradient approach should exist, and may be worth exploring in the future.  

\section{Acknowledgements}
This work was supported by the Deutsche Forschungsgemeinschaft (DFG, German Research Foundation) - CRC 1456/1 - 432680300.

\clearpage

\section{Supplement}
\linespread{1.3}

\renewcommand{\thefigure}{S\arabic{figure}}
\setcounter{figure}{0}
\renewcommand{\thetable}{S\arabic{table}}
\setcounter{table}{0}

\subsection{Computation of gradients}
The gradient of the log-likelihood was computed by the chain rule, first computing the derivatives with respect to each $I_{l,r,s}$. These derivatives are given by
\begin{equation*}
    \frac{\partial \log P(\mathcal I \,|\, \rho)}{\partial{I_{l,r,s}}} = \sum_{j = 1}^N \frac{1}{P(\vec k_1^{(j)},\dots,\vec k_{n_j}^{(j)} \,|\, \rho)}\frac{\partial P(\vec k_1^{(j)},\dots,\vec k_{n_j}^{(j)} \,|\, \rho)}{\partial{I_{l,r,s}}},
\end{equation*}
and computed by accumulation over the summands in eq.~\eqref{eq: likelihood  approx} of the main text.
To that end, we rewrite eq.~\eqref{eq: likelihood  approx} as 
\begin{equation*}
    P(\vec k_1^{(j)},\dots,\vec k_{n_j}^{(j)} \,|\, \rho) \approx \sum_{l = 1}^{n_\vec{R}} \sum_{s'=1}^{n_s} p_{ls'j},
\end{equation*}
where
\begin{equation*}
    p_{l,s',j} = w_l \exp(-I_0 \lambda_l) \frac1{n_s} \prod_{i=1}^{n_j} I_{l,r(\vec k_i^{(j)}),s(\vec k_i^{(j)})+s' \bmod {n_s}}.
\end{equation*}
Then 
\begin{equation*}
     \frac{\partial P(\vec k_1^{(j)},\dots,\vec k_{n_j}^{(j)} \,|\, \rho)}{\partial{I_{l,r,s}}} = \sum_{l = 1}^{n_\vec{R}} \sum_{s'=1}^{n_s} \frac{\partial p_{l,s',j}}{\partial{I_{l,r,s}}},
\end{equation*}
where $\frac{\partial p_{ls'j}}{\partial{I_{l,r,s}}}$ is obtained using the product rule as
\begin{align*}
     \frac{\partial p_{l,s',j}}{\partial{I_{l,r,s}}} = - a_{r,s} I_0 p_{l,s',j}
    + \sum_{i=1}^n \begin{cases}
        \frac{p_{l,s',j}}{I_{l,r,s}} & r = r(\vec k_i^{(j)}) \wedge s = s(\vec k_i^{(j)})+s' \bmod {n_s} \\
        0 & \text{else,}
    \end{cases}
\end{align*}
with the first summand taking into account the dependency of $\lambda_l$ on $I_{l,r,s}$ via eq.~\eqref{eq: numerical nphotons} of the main text. 
Finally, the derivatives with respect to the Gaussian bead positions $\vec y_i$ are obtained by the chain rule as
\begin{equation*}
    \frac{\partial}{\partial\vec y_i} \log P(\mathcal I \,|\, \rho) = \frac{\partial I_{l,r,s}}{\partial\vec y_i} \frac{\partial \log P(\mathcal I \,|\, \rho)}{\partial{I_{l,r,s}}}.
\end{equation*}
Taking the Fourier transform of eq.~\eqref{eq: representation}, $I_{l,r,s}$ is given by
\begin{align*}
    I_{l,r,s} = \lvert \hat\rho(\vec R_l \vec q_{r,s})\rvert^2 = \left\lvert\sum_{i=1}^m h_i \exp \left(-\frac{w_i^2 \lVert\vec q_{r,s}\rVert^2}{2}\right)\exp \left(\mathrm i (\vec R_l\vec q_{r,s}) \cdot \vec y_i \right) \right\rvert^2 
    % \\
    % &= \sum_{i=1}^m \sum_{j=1}^m h_i h_j \exp \left(-\frac{(w_i^2+w_j^2) \lVert\vec q_{r,s}\rVert^2}{2} + \mathrm i \vec q_{r,s}(\vec y_i-\vec y_j) \right)
\end{align*}
where $\vec q_{r,s}$ are the centers of the polar grid cells as defined in the main text. Note that the $\vec q_{r,s}$ are represented in Cartesian coordinates and merely arranged on a polar grid.
Finally, the gradient of $I_{l,r,s}$ is obtained as,
\begin{align*}
    \frac{\partial I_{l,r,s}}{\partial\vec y_i} &= 2\Re \left(\overline{\hat \rho(\vec q_{r,s})}\frac{\partial \hat \rho(\vec q_{r,s})}{\partial \vec y_i}\right) \\
    &= 2 h_i \vec R_l\vec q_{r,s} \exp \left(-\frac{w_i^2 \lVert\vec q_{r,s}\rVert^2}{2}\right) \Re \left( \mathrm i\overline{\hat \rho(\vec q_{r,s})}\exp \left(\mathrm i (\vec R_l\vec q_{r,s}) \cdot \vec y_i \right)\right).
\end{align*}
as we have derived in equations (55b), (59) and (60) of Ref.\ 40 of the main text.

\subsection{Definition of prior distribution}
The prior distribution $P(\rho_{\vec y_i})$ was given by 
\begin{equation}
    -\log P(\rho_{\vec y_i}) = \sum_{i=1}^m \sum_{j=1}^m f_1(\lVert \vec y_i - \vec y_j \rVert) + \sum_{i=1}^m f_3\left(\sum_{j=1}^m f_2(\lVert \vec y_i - \vec y_j \rVert)\right),
\end{equation}
where $f_1$, $f_2$ and $f_3$ are defined as follows in terms of the smoothstep function
\begin{equation}
    f(x, x_0, x_1, y_0, y_1) = y_0 + (y_1 - y_0) \,\tilde f\!\left(\frac{x - x_0}{x_1 - x_0}\right), \quad 
    \tilde f(x) = \begin{cases}
        0 & x < 0, \\
        3x^2-2x^3 & 0 \leq x \leq 1 \\
        1 & x > 1,
    \end{cases}
\end{equation}
which smoothly transitions between $y_0$ and $y_1$ on the interval from $x_0$ to $x_1$. 
The short range repulsive pair potential was given in terms of its derivative $f_1'(r) = f(r, 0.8, 0.9, s_1, 0) + s_2 \exp(-r^2/(2d(t)^2))$. 
The neighbor-counting function was given by $f_2(r) = f(r, 5, 10, 1, 0)$, and, finally $f_3$ was defined in terms of its derivative $f_3'(r) = f(r, 10, 20, s_3, 0)$. The Gaussian component of $f_1(r)$ was included to ensure a uniform distribution of the Gaussian beads at the beginning of the optimization. The parameter $d(t)$ was decreased to zero with increasing $t$ as shown in Figure S1. For the values $s_1$, $s_2$, and $s_3$ see Table S1.

\begin{table}[p]
    \begin{tabular}{llllllllll}
        \toprule
        Name & PDB & \shortstack[l]{batch\\size}  & \shortstack[l]{Lebedev\\order} & \shortstack[l]{radial\\grid dim.} & \shortstack[l]{angular\\grid dim.} & \shortstack[l]{optim.\\steps} & $s_1$ & $s_2$ & $s_3$ \\
        \midrule
        Crambin & 1EJG & $200\,000$ & $47$ & $60$& $64$ & $50\,000$ & $10$ & $30$ & $20$\\
        PDZ-domain & 1QAU & $50\,000$ & $65$ & $60$ & $96$ & $20\,000$ & $10$ & $10$ & $20$\\
        Lysozyme & 148L & $50\,000$ & $65$ & $120$ & $96$ & $20\,000$ & $10$ & $50$ & $20$\\
        \bottomrule
    \end{tabular}
    \caption{Parameters used for each optimization run. The orientations $\vec R_l$ from eq.~{\eqref{eq: likelihood approx}} where defined in terms of Lebedev grids of order $l$ as in our previous study \cite{schultze_bayesian_2024}. Grid points $a_{r,s}$ where uniformly spaced up to $\lVert a_{r,s}\rVert = 3\text{\AA}{}^{-1}$.}
    \label{tab: parameters supplement}
\end{table}

\begin{figure}[p]
    \centering
    \includegraphics[width=1\textwidth]{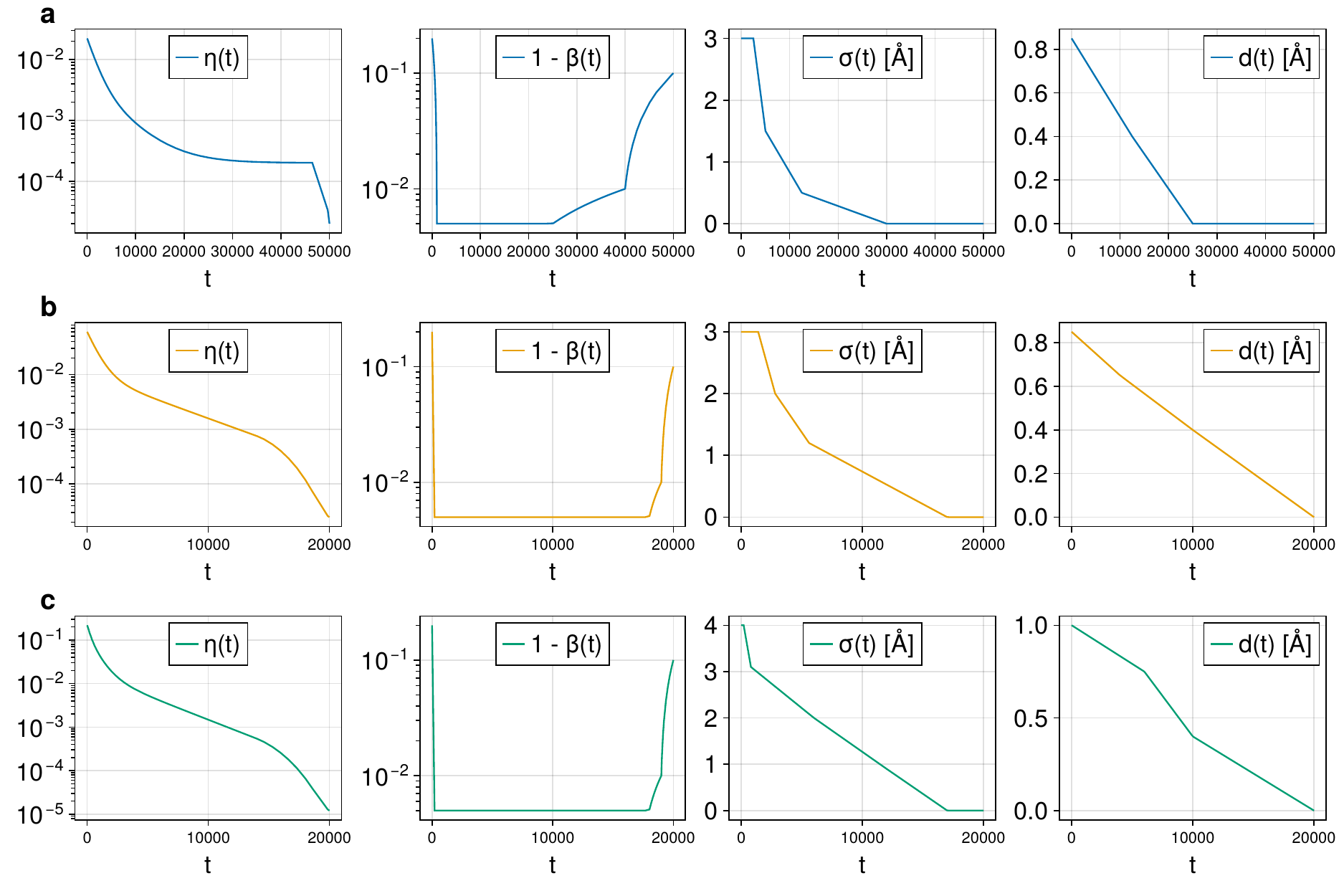}
    \caption{Time dependent parameters $\eta(t)$, $\beta(t)$, $\sigma(t)$, and $d(t)$ used in the optimization runs for \textbf{a} crambin, \textbf{b} PDZ-domain, and \textbf{c} Lysozyme.}
    \label{fig: parameters}
\end{figure}

\clearpage
\bibliography{xfel}

\end{document}